\documentclass[11pt]{article}
\usepackage{graphicx}       % for eps figures
\usepackage{bm}         % for bold math symbols
\usepackage{amsfonts}
\usepackage{amsmath}        % for \text and such
\usepackage{latexsym}

\newcommand{\el}{electric field}

        \newcommand{\ket}[1]{$ |{#1}\rangle$}
        
        \newcommand{\mket}[1]{ \left|{#1} \right\rangle }

\title{CNOT on Polarization States of Coherent Light}
\author{ Goce Chadzitaskos and Ji\v r\'{i} Tolar\\
Department of Physics, FNSPE, Czech Technical University in Prague,\\
B\v{r}ehov\'{a} 7, 115 19 Praha 1, Czech Republic. \\
E-mails: goce.chadzitaskos@fjfi.cvut.cz, jiri.tolar@fjfi.cvut.cz}

\begin{document}

\maketitle

\begin{abstract} We propose a CNOT gate for quantum
computation. The CNOT operation is based on existence of triactive
molecules, which in one direction have dipole moment and  cause
rotation of the polarization plane of linearly polarized light and
in perpendicular direction have a magnetic moment. The incoming
linearly polarized laser beam  is divided into two beams by beam
splitter. In one beam a control state is prepared  and the other
beam is a target. The interaction of polarized states of both beams
in a solution containing triactive molecules can be described  as
interaction of two qubits in CNOT.

\end{abstract}

%\keywords{quantum optics, nonlinear optics, quantum information}

\section{Introduction}

Numerous investigations have proposed various physical  ways of how
to realize an operating CNOT gate which is one of the basic elements
for quantum computing \cite{cnot1, cnot2}.  Our proposal is based on
the manipulation of polarization states of the laser beam which has
many advantages in technological aspects that are not necessary to
describe. The polarization state of completely polarized light can
be described by the same mathematics as a qubit. A geometric
representation of both quantities is provided by the Poincar\'e (or
the Bloch) sphere.

In order to be able to construct proposed  CNOT gate, triactive
molecules are needed as basic ingredient.  When polarized light
passes through a solution of an optically active compound --- chiral
or optical isomers or optical polymers --- the direction  of
polarization is rotated to the right or to the left. Recent
development in the chemistry of such dipole molecules with
controlled optical properties provides us with an opportunity to ask
chemists to prepare such an isomer or polymer molecule with one more
property --- with magnetic dipole moment oriented perpendicularly to
the electric dipole moment as shown in Fig.1. For instance, what we
need is to implement a magnetic dipole moment to some chiral isomer
or polymer, with high anisotropic polarizabilities as described in
\cite{Cdenz}. Thus we need a molecule with the following properties:

\begin{itemize}

\item  It has an electric dipole moment $ \overrightarrow{p}$

\item  It has a property of optical activity in one direction
$\overrightarrow{a}$, i.e., the direction of polarization  of
passing linearly polarized light rotates due to interaction with the
molecules whose $\overrightarrow{a}$ is oriented parallel with the
ray ($\overrightarrow{a}$ and $\overrightarrow{p}$ are often
parallel --- let us suppose  it in the following)

\item It has a magnetic moment $\overrightarrow{m}$ perpendicular to
 directions $\overrightarrow{a}$ and $\overrightarrow{p}$.

\end{itemize}

It may be expected that modern chemistry can produce such molecules,
since  molecules with the first two properties, i.e. without
magnetic moment, have already been produced. For molecules with
electric moment see \cite{Cdenz} Moreover, the engineering of
core-shell nanoparticles can be used to produce triactive molecules
\cite{Caruzo}.

\begin{figure}
  {\includegraphics[scale=0.6]{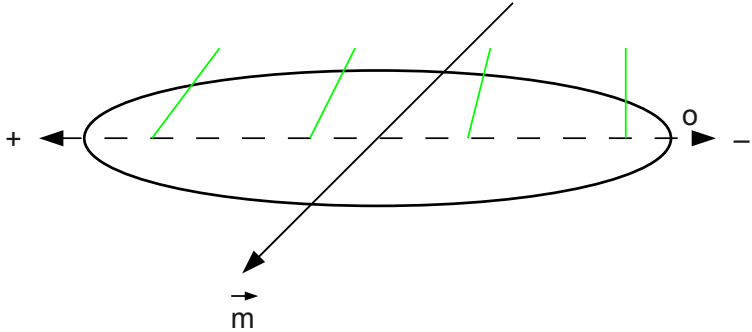}}
  \caption{\it Scheme of a triactive molecule. Molecule has
magnetic moment $\vec m$ in the direction perpendicular to the
electric dipole moment, and the direction of the electric dipole
moment (dashed line) is the axis of rotation of the direction of
polarization - green lines}
\end{figure}

\section{Operation}

For the sake of describing the principle of operation we will
consider completely linearly polarized light.

Let us have a  concentration $n$ of such triactive molecules in a
homogenous solution under thermal motion, they rotate in all
directions, the distribution of dipole moments of molecules is the
same in all directions and the polarization density is
$$\overrightarrow{P} = 0 .$$
The angle of rotation of  polarization  of passing light will depend
only on the path length of light in the solution.

Putting the solution in a strong homogenous magnetic field in an
ideal case guarantees that the thermal rotation of the molecules
will be restricted to precession of magnetic moments around the
direction of  the magnetic field. The magnetic field thus breaks the
isotropy of optical activity of solution. After switching the
magnetic field, the angles of rotation of polarization of light in
the directions parallel and perpendicular to the magnetic field at
the same distance are no more equal. The resulting angle of rotation
of polarization of passing light depends on the path length of light
in the solution and on the angle between the incident ray and the
magnetic field.

Next switching an \el in a direction perpendicular to the magnetic
field, the rotational symmetry in the plane perpendicular to the
magnetic field is broken. The polarization density will have the
direction of \el. The angle of rotation of polarization (passing the
same distance in solution) will be minimal in the direction of
magnetic field and maximal in the direction of \el.

Let us send two light beams through the solution in two
perpendicular directions, the first one being parallel with the
magnetic field and the second perpendicular.

The oscillating electric field of the light beam parallel with the
magnetic field controls the absolute value of polarization density
of the illuminated solution, and the polarization state of this beam
does not change (in an ideal case of low temperature). The
polarization of the light beam perpendicular to the magnetic field
will rotate around the direction of the first beam. Moreover, it can
also influence the mean absolute value of polarization density of
solution when the polarization of the second beam is not parallel
with the magnetic field.

The angle of  rotation of the polarization  of the beam
perpendicular to the magnetic field will depend on the direction of
the polarization density of the solution, which will lie in the
plane perpendicular to the magnetic field. Of course, the dependence
of the angle of rotation of the polarization on the temperature as
well as the length of the path of beam in the solution have to be
taken into consideration, and these parameters will guarantee the
correct function of CNOT gate.

\section{How does it work?}

The operation of the proposed CNOT gate can  be described as
follows:
\begin{enumerate}
\item The linearly polarized laser beam (LB) is divided into two
parts by the  beam splitter (BS), see Fig. 2.
\item The optical devices C and T in
the two  branches prepare the polarization states according to the
requirement for the input states in CNOT cell.
\item One of divided
beams which passes the CNOT cell with the solution of three-active
molecules parallel with the external magnetic field plays the role
of the control qubit  and the other is the target qubit. Both beams
pass the CNOT cell simultaneously.
\end{enumerate}
The control beam determines the magnitude of the absolute value of
the polarization density in cube and the direction of the
polarization of passing light. According to the Langevin--Debye
theory the mean value of the polarization density is
$$ <|P|> =   \frac{n p^2 \pi }{\sqrt{2} k T} E_{C} ,$$
where $\vec{E_C}$  is an amplitude of the vector of the intensity of
electric field of control beam, and  the direction of the
polarization density is given by the direction of linearly polarized
light in the control channel. The influence of the vector of the
target electric field  is negligible because it rotates several
times when passing the CNOT cube.

The nonzero mean value of the magnitude of the vector of
polarization density in the described direction is the crucial point
for the operation of the CNOT. If the direction of polarization of
the control beam will be parallel with the direction of target beam,
the polarization of target beam will rotate more than in the case
where the direction of polarization of the control beam will be
perpendicular to the direction of target beam. The direction of
polarization of the control beam is not changed because the magnetic
field does not allow significant changes of the polarization density
of solution in direction of control beam. So one can control the
geometry of the cube, concentration of triactive molecules in the
solution and temperature to fit the difference between the
polarization planes of the target beam of two previous cases after
passing the cube.

\begin{figure}
  {\includegraphics[scale=0.6]{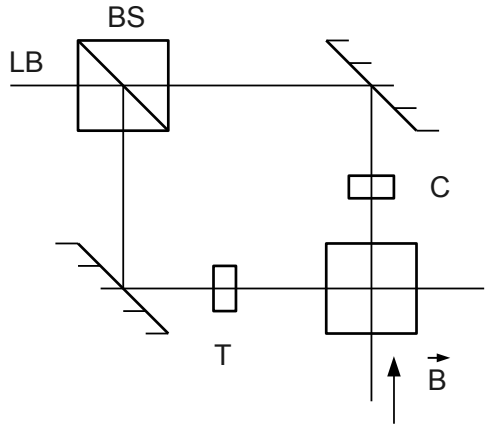}}
  \caption{\it Scheme of CNOT in operation.
The laser beam ({\bf LB}) is divided by the beam splitter ({\bf BS})
into control and target parts. Control and target states are
prepared using C and T which manipulate the polarization in each
state. The control and target beams then enter the CNOT cell.
Control part passes through the cell in the direction of external
magnetic field $\vec{\bf B}$ and target part in perpendicular
direction.}
\end{figure}

The operation is shown in Fig.2. Let \ket{0} be polarization state
of linearly polarized light perpendicular to the plane of figure --
vertical polarization, and \ket{1} in the plane -- horizontal
polarization. Having the input state \ket{1}, after splitting the
state is $\mket{1} \otimes \mket{1}$. The device C change the state
\ket{1} into $\mket{\gamma}$ and T change the state \ket{1} into
$\mket{\tau}$. The state on inputs in cube is $\mket{\gamma} \otimes
\mket{\tau}$. On the output of cube the state is $\mket{\gamma}
\otimes \mket{\tau - \gamma}$.

\begin{tabular}{|c|c|c|c|}
\hline \multicolumn{4}{|c|}{Table: CNOT operations on polarization states of beams} \\
\hline  Input control & Input target & Output control & Output target \\
\hline $\rightarrow$ & $\rightarrow$ & $\rightarrow$ & $\rightarrow$ \\
\hline $\rightarrow$ & $\uparrow$ & $\rightarrow$ & $\uparrow $\\
\hline $\uparrow$ & $\rightarrow$ & $\uparrow$ & $\uparrow$ \\
\hline $\uparrow$ & $\uparrow$ & $\uparrow$ & $\rightarrow$ \\
\hline
\end{tabular}

\section{Conclusion}

The magnetic dipole moment is in principle necessary in order to fix
an axis of rotation of molecules. The same effect can be achieved
also by other means, e.g. ordering of optically active dipole
molecules in material with a preferred plane of change in
polarizability. We note that in spite of the fact that the
interactions involved will be very weak (perhaps of the order of
$\chi^{(3)}$ effects) the proposed mechanism is worthy of further
investigation.

\section*{Acknowledgements}
The support by the Ministry of Education of Czech Republic (projects
MSM6840770039 and LC06002) is acknowledged. The authors are grateful
to the referee whose comments helped to improve the presentation.

\end{document}